\documentclass[onecolumn, preprintnumbers, superscriptaddress, aps, prl, 12pt, floatfix, tightenlines]{revtex4-1}               

\usepackage{CJKutf8}
\usepackage{latexsym}
\usepackage{amstext}
\usepackage{amsfonts}
\usepackage{dsfont}
\usepackage{color}
\usepackage{amssymb}
\usepackage{amsmath}
\usepackage{relsize}
\usepackage{amsxtra}
\usepackage{verbatim}
\usepackage{hyperref}
\usepackage{graphicx}

\AtBeginDvi{\input{zhwinfonts}}
\begin{document}
\begin{CJK*}{UTF8}{zhkai}

\title{Distinct itinerant spin-density waves and local-moment \\ 
antiferromagnetism in an intermetallic ErPd$_2$Si$_2$ single crystal}

\author{Hai-Feng Li} 
\email{hfli@ill.fr; hfli2101@gmail.com}
\affiliation{J$\ddot{u}$lich Centre for Neutron Science JCNS, Forschungszentrum J$\ddot{u}$lich GmbH, Outstation at Institut Laue-Langevin, Bo$\hat{\imath}$te Postale 156, F-38042 Grenoble Cedex 9, France}
\affiliation{Institut f$\ddot{u}$r Kristallographie der RWTH Aachen University, D-52056 Aachen, Germany}
\author{Chongde Cao}
\affiliation{Key Laboratory of Space Applied Physics and Chemistry of Ministry of Education, Department of Applied Physics, Northwestern Polytechnical University, Xi’an 710072, P.R. China}
\author{Andrew Wildes}
\affiliation{Institut Laue-Langevin, Bo$\hat{\imath}$te Postale 156, F-38042 Grenoble Cedex 9, France}
\author{Wolfgang Schmidt}
\affiliation{J$\ddot{u}$lich Centre for Neutron Science JCNS, Forschungszentrum J$\ddot{u}$lich GmbH, Outstation at Institut Laue-Langevin,
Bo$\hat{\imath}$te Postale 156, F-38042 Grenoble Cedex 9, France}
\author{Karin Schmalzl}
\affiliation{J$\ddot{u}$lich Centre for Neutron Science JCNS, Forschungszentrum J$\ddot{u}$lich GmbH, Outstation at Institut Laue-Langevin,
Bo$\hat{\imath}$te Postale 156, F-38042 Grenoble Cedex 9, France}
\author{Binyang Hou}
\affiliation{European Synchrotron Radiation Facility, Bo$\hat{\imath}$te Postale 220, F-38043 Grenoble Cedex, France}
\author{Louis-Pierre Regnault}
\affiliation{SPSMS, UMR-E 9001, CEA-INAC/UJF-Grenoble 1, MDN, 17 rue des Martyrs, F-38054 Grenoble Cedex 9, France}
\author{Cong Zhang}
\affiliation{Institut f$\ddot{u}$r Kristallographie der RWTH Aachen University, D-52056 Aachen, Germany}
\author{Paul Meuffels}
\affiliation{Peter Gr$\ddot{u}$nberg Institut PGI and JARA-FIT, Forschungszentrum J$\ddot{u}$lich GmbH, D-52425 J$\ddot{u}$lich, Germany}
\author{Wolfgang L$\ddot{\texttt{o}}$ser}
\affiliation{Leibniz-Institut f$\ddot{u}$r Festk$\ddot{o}$rper- und Werkstoffforschung (IFW) Dresden, Postfach 270116, D-01171 Dresden, Germany}
\author{Georg Roth}
\affiliation{Institut f$\ddot{u}$r Kristallographie der RWTH Aachen University, D-52056 Aachen, Germany}

\date{\today}

\begin{abstract}

\textbf{Identifying the nature of magnetism, itinerant or localized, remains a major challenge in condensed-matter science. Purely localized moments appear only in magnetic insulators, whereas itinerant moments more or less co-exist with localized moments in metallic compounds such as the doped-cuprate or the iron-based superconductors, hampering a thorough understanding of the role of magnetism in phenomena like superconductivity or magnetoresistance. Here we distinguish two antiferromagnetic modulations with respective propagation wave vectors of \textbf{Q}$_{\pm}$ = ($H \pm 0.557(1)$, 0, $L \pm 0.150(1)$) and \textbf{Q}$_\texttt{C}$ = ($H \pm 0.564(1)$, 0, $L$), where $\left(H, L\right)$ are allowed Miller indices, in an ErPd$_2$Si$_2$ single crystal by neutron scattering and establish their respective temperature- and field-dependent phase diagrams. The modulations can co-exist but also compete depending on temperature or applied field strength. They couple differently with the underlying lattice albeit with associated moments in a common direction. The \textbf{Q}$_{\pm}$ modulation may be attributed to localized 4\emph{f} moments while the \textbf{Q}$_\texttt{C}$ correlates well with itinerant conduction bands, supported by our transport studies. Hence, ErPd$_2$Si$_2$ represents a new model compound that displays clearly-separated itinerant and localized moments, substantiating early theoretical predictions and providing a unique platform allowing the study of itinerant electron behavior in a localized antiferromagnetic matrix. \\}

\end{abstract}

\maketitle
\end{CJK*}


Unravelling the interplay between opposite but also complementary phenomena stands at the forefront of condensed-matter science. For example, unconventional Cooper pairs, on the one hand, can be glued by the common thread of spin fluctuations; on the other hand, they can be competitively ruined by the formation of long-ranged ferromagnetic (FM) or antiferromagnetic (AFM) ordering \cite{Yosida1996, Vojta2009, Schmitt2012, Kou2009, Scalapino2012, Kamihara2008, LiLFAO, Zhao2009, LiBC122, Tucker2012, Steffens2013, Zhang2013, Steglich1979}. The archetypal picture of magnetism displays an opposing dual character, i.e., itinerant electron magnetism with weak interactions and localized moments with strong Coulomb repulsions \cite{Yosida1996}. Understanding the behaviour of itinerant electrons in the presence of localized moments may shed light on nontrivial properties of correlated electron systems such as spin- or charge-density waves or superconductivity \cite{Yosida1996, Vojta2009, Schmitt2012, Kou2009, Scalapino2012, Kamihara2008, LiLFAO, Zhao2009, LiBC122, Tucker2012, Steffens2013, Zhang2013, Steglich1979, Jin1994, Diallo2009, Diallo2010, Ewings2011, Ke2011, Hu2012, Vilmercati2012, You2013, Wu2011, Fawcett1988, Gruener1988, Gruener1994, Yin2014}, for which uniting both types of magnetism in one compound with clear electronic origins is necessary \cite{Wu2011, Brown1994}. Experimentally, it is hard to clearly distinguish itinerant from localized moments in 3\emph{d}-based strongly-correlated electron materials like the copper-oxide superconductors \cite{Vojta2009, Wu2011}, despite the fact that they can be theoretically modeled \cite{Yosida1996}. Consequently, the experimentally observed smaller moment size in point compared to the expected theoretical saturation value \cite{Lacroix2011} can be attributed either to the screening effect of itinerant electrons or to the frustration effect of localized spins. In addition, the nature of the antiferromagnetism or spin-density waves (SDWs) of iron-superconductors is still hotly debated as to whether the magnetic neutron excitations can be best described by the itinerant or localized picture \cite{Diallo2009, Zhao2009, Diallo2010, Ewings2011, Ke2011, Hu2012, Vilmercati2012, You2013}. Disentangling these arguments necessitates the search for a model system that hosts clearly-defined itinerant and localized spins, thus permitting a complete understanding of their coupling mechanism.

The 5\emph{f} electrons in actinides such as U-based compounds UPt$_3$ and UPd$_2$Al$_3$ show experimental and theoretical evidence for a localized and delocalized dual nature \cite{Schoenes1996, Bernhoeft1998, Zwicknagl2003, Durakiewicz2006, Zwicknagl2006, Efremov2004, Troc2012, Klimczuk2012, Magnani2007} which may play an important role in producing heavy-fermion superconductivity \cite{Sato2001}. In strongly spin-orbit coupled systems such as 4\emph{f}- or 5\emph{f}-based compounds, a novel spin-orbit density wave was proposed as an emergent quantum phase with a breaking of translational while preserving time-reversal symmetry, which theoretically sheds light on the intriguing {\lq\lq}hidden-order{\rq\rq} of URu$_2$Si$_2$ \cite{Das2012}. For lanthanide-based compounds, the rare-earth (\emph{RE}, except cerium) 4\emph{f} moments are generally localized because unpaired 4\emph{f} electrons are well shielded by the $5s^2p^6$ shells. The $4f^n5d^m$6$s^2$ (\emph{n} = 2-7 and 9-14; \emph{m} = 0, 1) valence electrons in lanthanide-based conductors act usually as a mediator for the interactions between 4\emph{f} moments in Ruderman-Kittel-Kasuya-Yosida (RKKY) exchanges (i.e., the long-range ordered 4\emph{f} moments interact indirectly with each other via conduction bands since the direct coupling between localized 4\emph{f} moments is generally weak) \cite{Jensen1991}, or are ferromagnetically polarized into magnetic polarons (i.e., local short-range FM regimes) by the localized 4\emph{f} moments \cite{Heikes1964, Moln2007, Li2012}. It is thus difficult for the conduction electrons to form a long-ranged AFM ordering. However, taking into account the coupling between nesting electrons and hole parts of Fermi surface, the 4\emph{f}-based conductors could provide the possibility for combining localized 4\emph{f} moments and SDWs of conducting electrons.

The SDW state, a low-energy \emph{self}-\emph{organized} collective modulation of electron spins, often appears in electronic conducting materials such as organic linear-chain compounds, low-dimensional metals or superconductors \cite{Fawcett1988, Gruener1988, Steglich1979, Gruener1994, Brown1994}. Since their first observation in chromium \cite{Fawcett1988}, SDWs display progressively appealing low-temperature properties, e.g., a proximity with charge-density waves (CDWs) and unconventional superconductivity \cite{Fawcett1988, Gruener1988, Steglich1979, Gruener1994, Brown1994, Wu2011, Johannes2008}. Above a threshold field, SDWs can be described as a set of delocalized AFM spins \cite{Gruener1994}.

\begin{figure*}[!ht]
\centering \includegraphics[width = 0.78\textwidth] {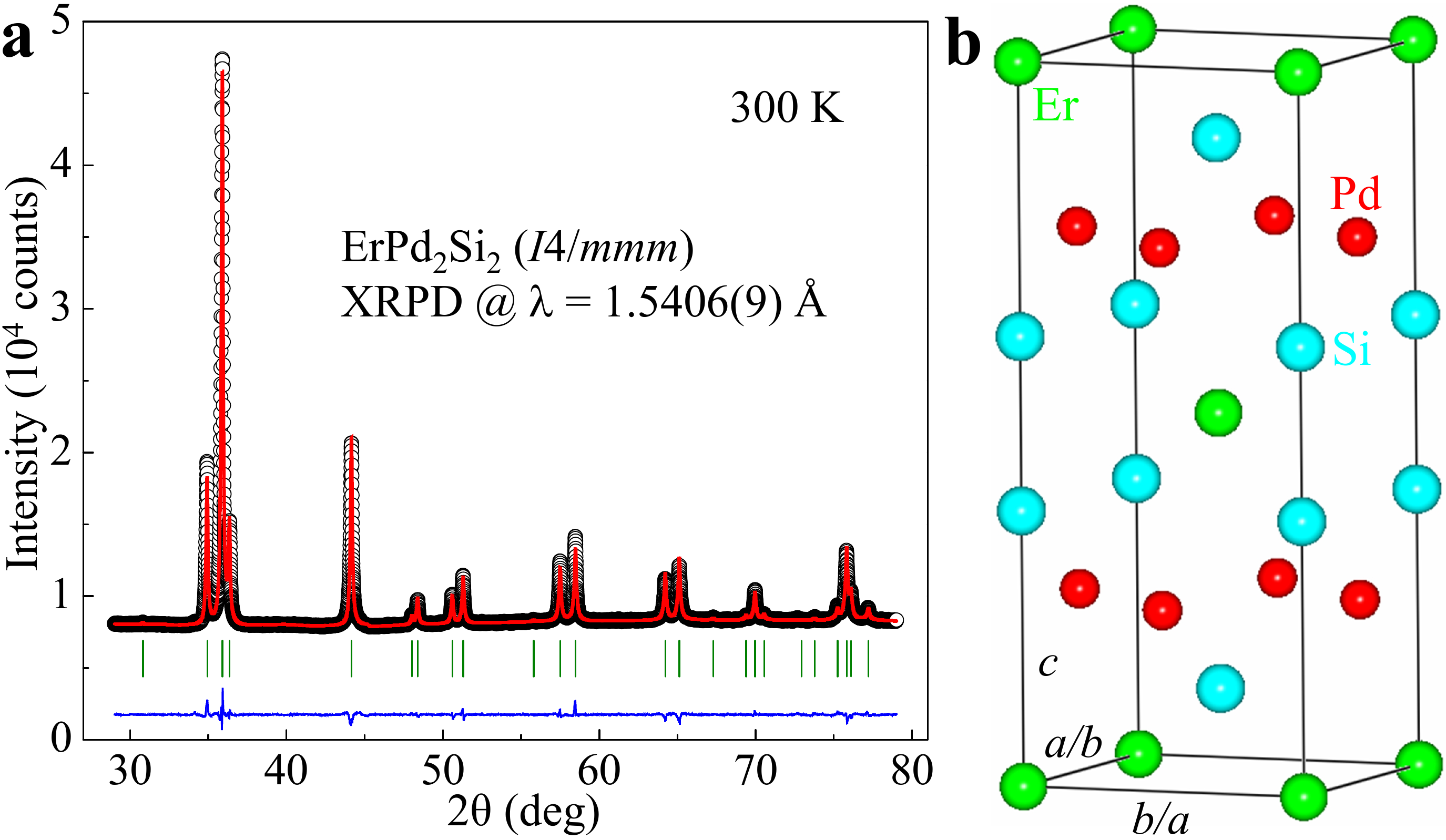}
\caption{
\textbf{Powder X-ray diffraction data and crystal structure of single-crystal ErPd$_2$Si$_2$ at 300 K.}
(a) Observed (circles) and calculated (solid lines) in-house X-ray powder-diffraction (XRPD) patterns of a pulverized ErPd$_2$Si$_2$ single crystal at ambient conditions obtained on an in-house diffractometer employing the copper $K_{\alpha1}$ = 1.5406(9) {\AA} radiation. The vertical bars mark the positions of nuclear Bragg reflections, and the lower curves represent the difference between observed and calculated patterns. Within the experimental accuracy, no detectable impurity phase is present. The results of the refinement are listed in Table 1. (b) The corresponding room-temperature crystal structure as refined.
}
\label{Fig1}
\end{figure*}

Intermetallic \emph{RE}Pd$_2$Si$_2$ silicides all crystallize with the same ThCr$_2$Si$_2$-type (Fig.~\ref{Fig1}) tetragonal $I4/mmm$ structure (\emph{a} = \emph{b} = 4.0987(1) {\AA}, and \emph{c} = 9.8762(1) {\AA} at ambient conditions, as listed in Table 1) as that of the family of 122-iron-pnictides \cite{Sasmal2008, LiSr122, Lee2010} and exhibit a wide range of interesting physical properties, e.g., pressure-induced superconductivity in CePd$_2$Si$_2$ and anomalous valence fluctuations in EuPd$_2$Si$_2$ \cite{Stewart2001}. Early theoretical proposals \cite{Kulikov1987} for the 4\emph{f} conductors with extended RKKY-interactions predicted that localized 4\emph{f} moments can promote a SDW state in the itinerant conduction electrons. So far, to our knowledge, no clear example of such a material with \emph{distinguishable} propagation vectors has been identified. Here we report on the first single-crystal neutron scattering study of ErPd$_2$Si$_2$ \cite{Tomala1994, Bazela1994, Sanchez1996, Venturini1996, Sampa2008, Cao2013}. We discover two distinct incommensurate spin states and attribute one to the localized 4\emph{f} electrons while attributing the other mainly to itinerant valence bands. We also build a detailed knowledge of the virtual coupling between both states, which is actually intractable in 3\emph{d}-metallic systems. Our findings correspond to theoretical predictions \cite{Kulikov1987} and thus establish a new model material.

\textbf{Results}

\begin{figure*}[!ht]
\centering \includegraphics[width = 0.78\textwidth] {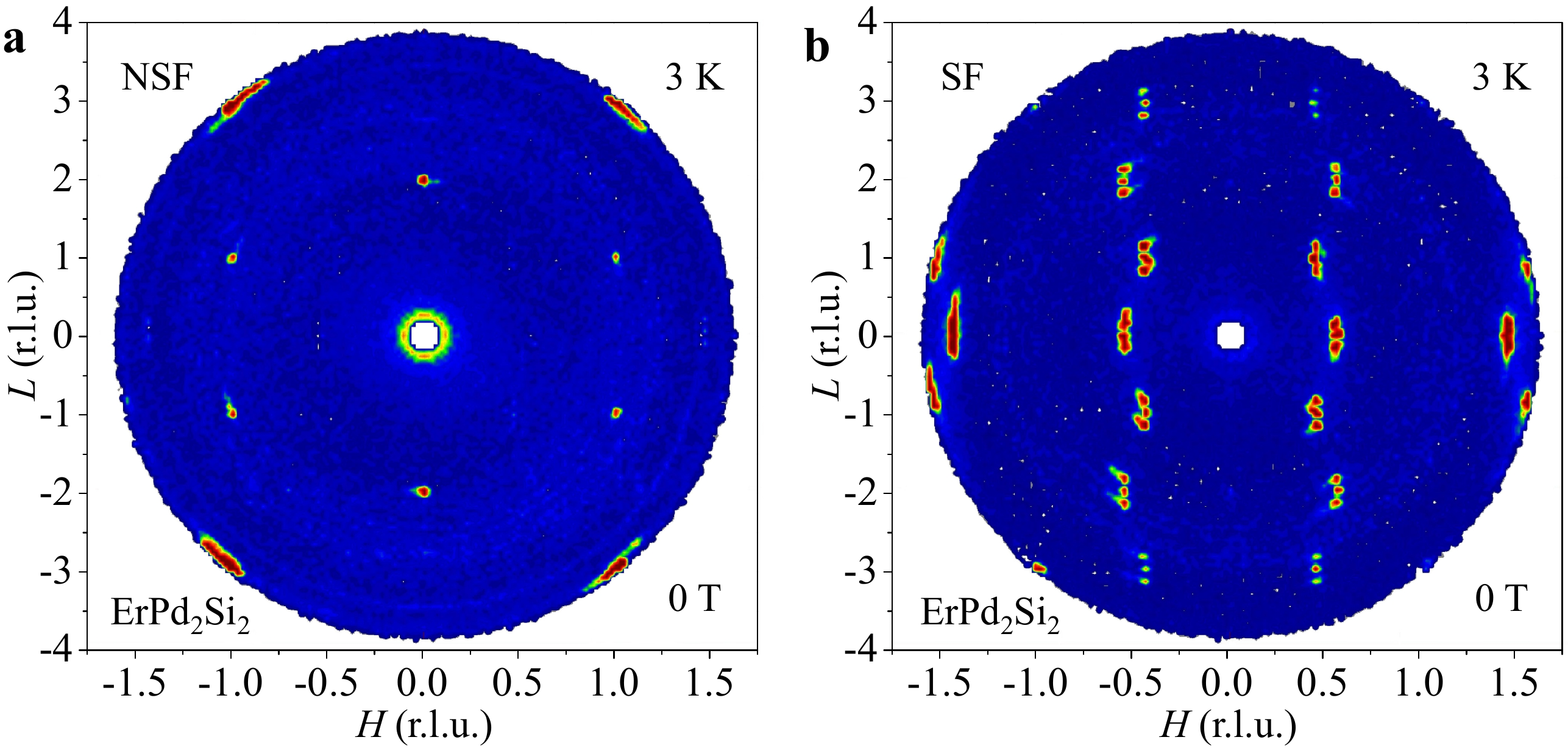}
\caption{
\textbf{Neutron polarization analysis data measured at 3 K using D7 (ILL).}
(a) The non-spin-flip, i.e., NSF, channel. This channel measures nuclear scattering and magnetic scattering from moment components parallel to $\hat{\textbf{P}}$. (b) The spin-flip, i.e., SF, channel. This channel measures magnetic scattering from moment components lying in the scattering plane. Here the same colour code is used for both intensities. In this study, the neutron polarization, $\hat{\textbf{P}}$, is perpendicular to the scattering plane (\emph{H}, 0, \emph{L}).
}
\label{Fig2}
\end{figure*}

\textbf{Neutron diffraction with polarization analysis.} Figure~\ref{Fig2} shows data from neutron diffraction with polarization analysis after the sample was cooled to 3 K in zero field.  Nuclear coherent scattering of polarized neutrons will not flip the neutron spin.  Magnetic scattering via polarized neutrons can flip the neutron spin, and is determined by the relative direction of the neutron polarization vector $\hat{\textbf{P}}$ with regard to both the scattering vector $\hat{\textbf{Q}}$ and the direction of the ordered-moments ${\hat{\boldsymbol{\mu}}}$ \cite{Stewart2009, Li2014Tb}. In our study, $\hat{\textbf{P}}$ is normal to the scattering plane (\emph{H}, 0, \emph{L}), i.e., parallel to the \emph{b} (or \emph{a}) axis. In this case, the moment-dependent cross-sections can be written as:
\begin{eqnarray}
\label{pol1_aw}
&& (\frac{d\sigma}{d\Omega})_{_{\rm SF}} = \frac{2}{3}(\frac{d\sigma}{d\Omega})_{_{\rm nsi}} +
(\frac{d\sigma}{d\Omega})^{^{\perp}}_{_{\rm mag}}, \texttt{and}                                                        \\
\label{pol2_aw}
&& (\frac{d\sigma}{d\Omega})_{_{\rm NSF}} = \frac{1}{3}(\frac{d\sigma}{d\Omega})_{_{\rm nsi}} + (\frac{d\sigma}{d\Omega})_{_{\rm nuc}} + (\frac{d\sigma}{d\Omega})^{^{\|}}_{_{\rm mag}} \texttt{,}
\end{eqnarray}
where the subscripts NSF and SF refer to the non-spin flip (NSF) and spin flip (SF) cross-sections, \emph{nsi} means the nuclear spin incoherent contribution, \emph{nuc} refers to nuclear coherent and isotopic incoherent contributions.  The magnetic contributions, labelled by \emph{mag}, are from two parts: the component of ${\hat{\boldsymbol{\mu}}}$ that is along the direction of $\hat{\textbf{P}}$ contributes to NSF scattering, i.e.,
\begin{eqnarray}
\label{pol4}
&& (\frac{d\sigma}{d\Omega})^{^{\|}}_{_{\rm mag}} {\propto} {\langle}{\hat{\mu}} \parallel \hat{\textbf{P}}{\rangle}^2;
\end{eqnarray}
the component that is normal to both $\hat{\textbf{P}}$ and $\hat{\textbf{Q}}$ gives rise to SF scattering \cite{Stewart2009}, i.e.,
\begin{eqnarray}
\label{pol3}
&& (\frac{d\sigma}{d\Omega})^{^{\perp}}_{_{\rm mag}} {\propto} {\langle}{\hat{\mu}} \perp \left(\hat{\textbf{P}} \times \hat{\textbf{Q}}\right){\rangle}^2.
\end{eqnarray}

\begin{figure*}[!ht]
\centering \includegraphics[width = 0.78\textwidth] {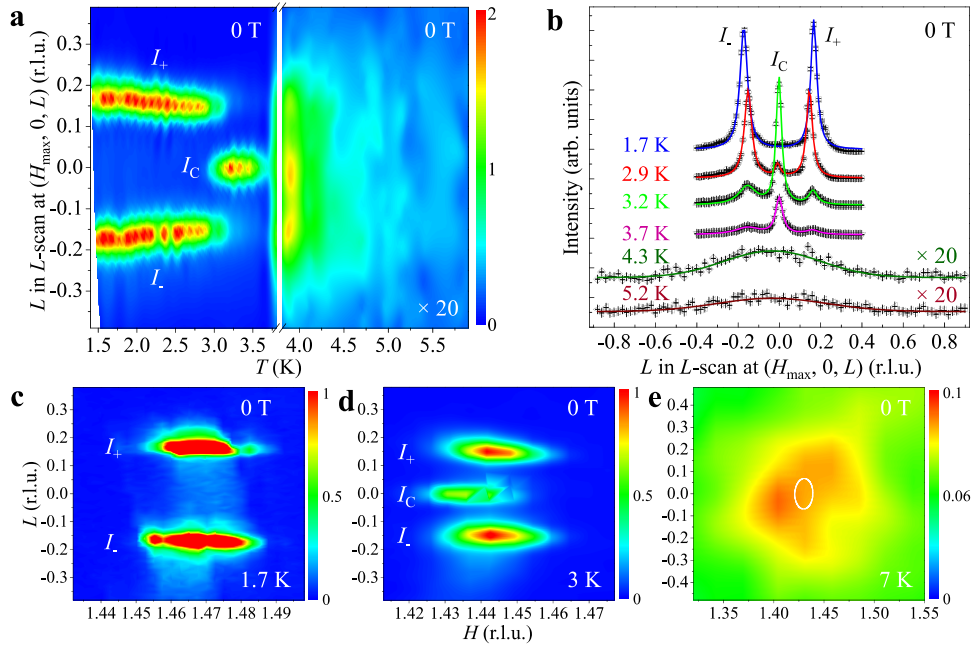}
\caption{
\textbf{Temperature dependence and reciprocal space maps in zero field using IN12 (ILL).}
(a) Temperature dependent \emph{L}-scans at ($H_{\texttt{max}}$, 0, \emph{L}). To correctly monitor the \emph{L}-scans we first located the $H_{\texttt{max}}$ for $I_\texttt{C}$ at each temperature. (b) \emph{L}-scans at representative temperatures. The intensity is vertically shifted for clarity. The solid lines are Lorentzian fits. Error bars are statistical errors. To clearly display the existence of the magnetic diffuse scattering as shown in (a) and (b), the observed magnetic intensity above 3.8 K is multiplied by 20. (c-e) \textbf{Q}-map in the (\emph{H}, 0, \emph{L}) scattering plane at 1.7, 3 and 7 K, respectively. The ellipse in (e) represents the expected \textbf{Q}-resolution.
}
\label{Fig3}
\end{figure*}

The NSF channel shows only the nuclear Bragg reflections structurally permitted by the tetragonal space group. The SF scattering shows magnetic Bragg peaks with two incommensurable propagation vectors. One set of peaks, that will be called $I_+$ and $I_-$ (Fig.~\ref{Fig3}), appear at \textbf{Q}$_{\pm}$ = ($H \pm 0.557(1)$, 0, $L \pm 0.150(1)$) and the second set, which will be called $I_\texttt{C}$ (Fig.~\ref{Fig3}), appear at \textbf{Q}$_\texttt{C}$ = ($H \pm 0.564(1)$, 0, $L$), where $\left(H, L\right)$ are the Miller indices for allowed nuclear reflections.

In our study, the $a$ and $b$ axes are equivalent in the tetragonal symmetry.  There is no magnetic signal in the NSF channel (Fig.~\ref{Fig2}a), therefore, the moments lie in the $\left(H, L\right)$ plane, mainly along the crystallographic $c$ axis \cite{Stewart2009} consistent with the powder neutron-diffraction study \cite{Tomala1994}. The uniaxial moment direction determined for the two spin states may be attributed to the Ising property of Er ions with a dominant crystal-field ground state {$|$}{$\pm 15/2$}{$>$} which leads to a high magnetic anisotropy along the \emph{c} axis \cite{Tomala1994}. The observed propagation vectors differ from those proposed in the neutron-powder-diffraction studies \cite{Tomala1994, Bazela1994} where positive and negative momenta cannot technically be differentiated.

\begin{figure*}[!ht]
\centering \includegraphics[width = 0.78\textwidth] {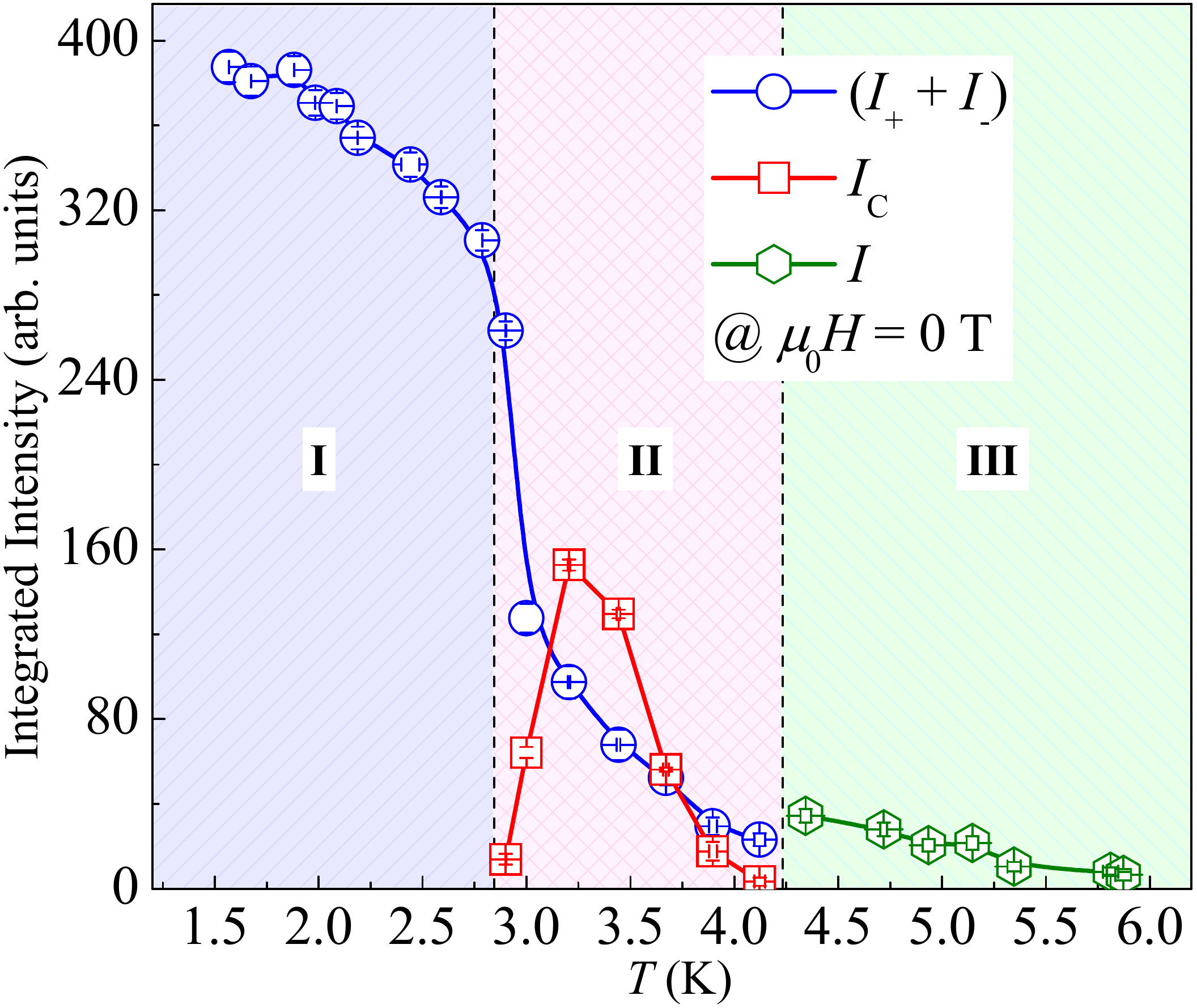}
\caption{
\textbf{Temperature-dependent phase diagram of the two spin states in zero field.}
Integrated magnetic intensity from the \textbf{Q}-scans versus temperature. The phase diagram clearly shows three regimes. We attribute regime I to the AFM state from purely localized 4\emph{f} moments; regime II to a mixture of localized 4\emph{f} moments with a SDW from weakly-pinned collective spins in the valence bands (details in the text); and regime III to large amplitude short-ranged spin orders. The error bars are from Lorentzian fits.
}
\label{Fig4}
\end{figure*}

\textbf{Reciprocal space maps and temperature-dependent phase diagram.} Figure~\ref{Fig3} shows the temperature variation of the magnetic Bragg peaks. Only the $I_\pm$ peaks are visible at the lowest temperature 1.7 K (Fig.~\ref{Fig3}c) in this study. As shown in Figs.~\ref{Fig3}a and b, when the temperature is increased there is a slight decrease in the peak intensities and the fractional $L$.  The intensities drop sharply at $\sim$ 2.9 K , and the $I_\texttt{C}$ peak suddenly appears (Fig.~\ref{Fig3}d). The intensities for $I_\pm$ continue to decrease gradually with increasing temperature while $I_\texttt{C}$ intensity rises sharply to reach a maximum at $\sim$ 3.2 K before rapidly decreasing. The peaks cease to be sharp resolvable features and combine to become a weak diffuse peak above $T_\texttt{N} \sim$ 4.2 K (Fig.~\ref{Fig3}e). The temperature-dependent intensities are summarized in Fig.~\ref{Fig4}, showing a phase diagram that may be divided into three clear regimes.

\begin{figure*}[!ht]
\centering \includegraphics[width = 0.78\textwidth] {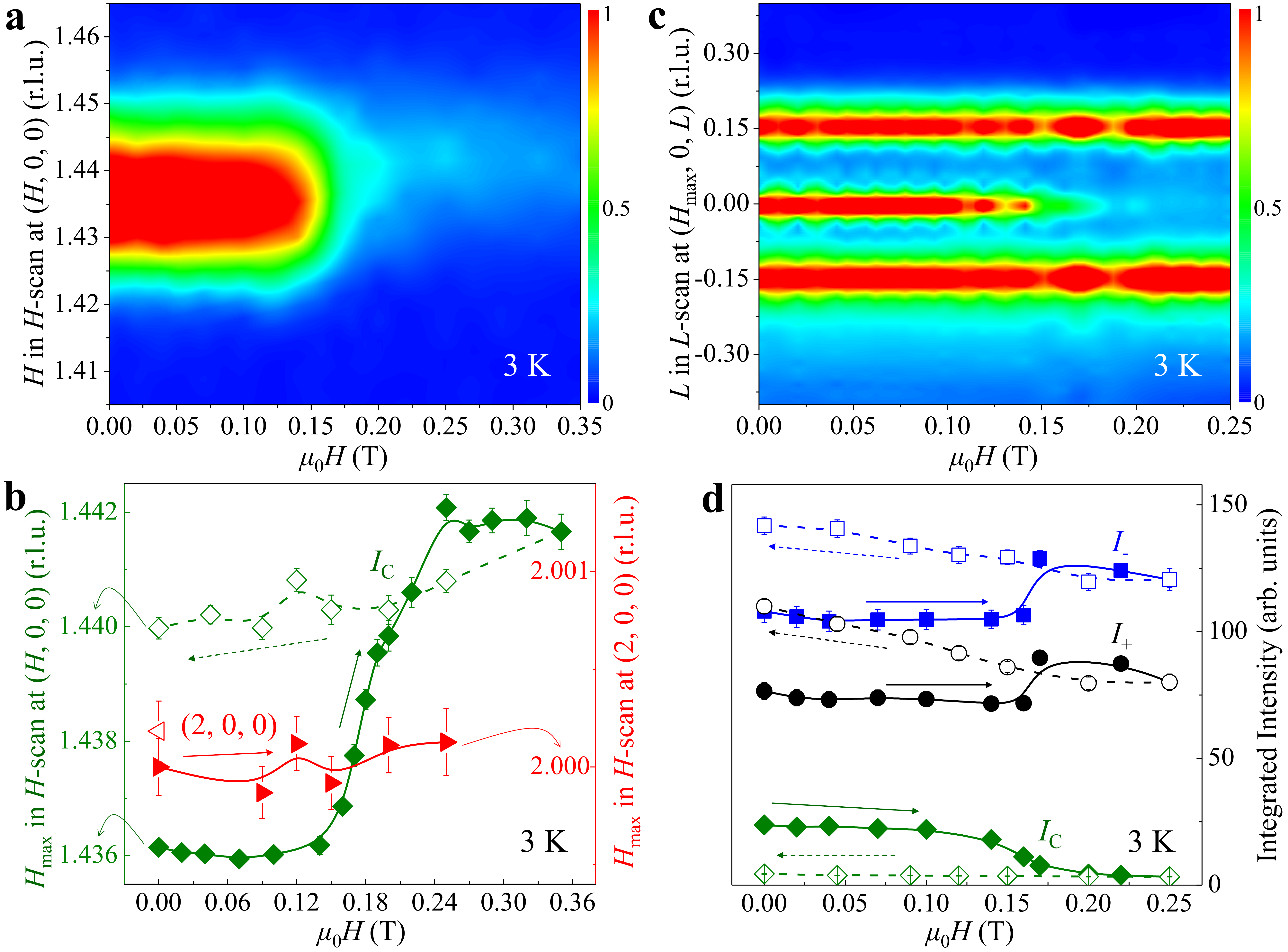}
\caption{
\textbf{Magnetic-field dependent phase diagram of the two spin states at 3 K.}
(a) \emph{H}-scans across the $I_\texttt{C}$ peak with increasing magnetic field. (b) $H_{\texttt{max}}$ (obtained from our Lorentzian fits) of the \emph{H}-scans around $I_\texttt{C}$ (diamonds) and nuclear (2, 0, 0) (triangles) Bragg peaks as a function of both increasing (solid symbols) and decreasing (void symbols) magnetic field (directions as marked). (c) \emph{L}-scans at ($H_{\texttt{max}}$, 0, \emph{L}) with increasing magnetic field, with $H_{\texttt{max}}$ taken from (b). (d) Integrated magnetic intensities of the $I_{\texttt{+}}$ (circles) and $I_{\texttt{-}}$ (squares) peaks (along the \emph{L}-direction) and of the $I_{\texttt{C}}$ (diamonds) peaks (along the \emph{H}-direction) as a function of increasing (solid symbols) and decreasing (void symbols) magnetic field. The lines are guides to the eye. The error bars are from Lorentzian fits.
}
\label{Fig5}
\end{figure*}

\begin{figure*}[!ht]
\centering \includegraphics[width = 0.78\textwidth] {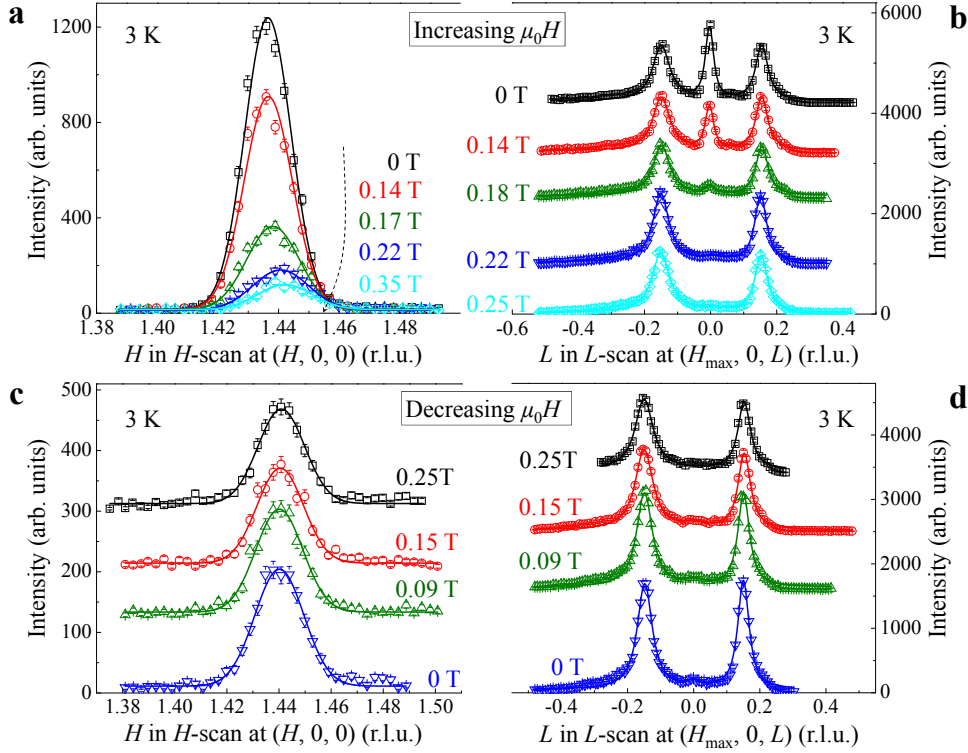}
\caption{
\textbf{Q-scans around the two spin states at representative fields as marked and 3 K.}
(a) \emph{H}-scans around (\emph{H}, 0, 0) at selected fields clearly showing that the \emph{H}-center shifts to higher values as the field increases. (b) \emph{L}-scans around ($H_{\texttt{max}}$, 0, 0) with $H_{\texttt{max}}$ taken from (a) at selected fields clearly showing that with increasing field, the central peak corresponding to the SDW state weakens largely and almost vanishes into the background above $\sim$ 0.18 T. (c) \emph{H}-scans around (\emph{H}, 0, 0) at selected fields clearly showing that with decreasing field, the \emph{H}-center keeps almost unchanged. (d) \emph{L}-scans around ($H_{\texttt{max}}$, 0, 0) at selected fields clearly showing that while decreasing the field to 0 T, the central peak corresponding to the SDW state does not come back as shown in (b). These interesting field- and temperature-dependent behaviors indicate that the itinerant SDW state, unlike that of the conventional long-range AFM ordering (a magnetic domain), is much more delicate consistent with the previous M$\ddot{\texttt{o}}$ssbauer study \cite{Tomala1994}. The solid lines in (a-d) are Lorentzian fits, and error bars are statistical errors. For clarity, the intensities in (b-d) are vertically shifted.
}
\label{Fig6}
\end{figure*}

\textbf{Magnetic-field dependent phase diagram.} The magnetic-field dependence of the intensities were measured at 3 K, which is the point where the intensities of $I_\pm$ and $I_\texttt{C}$ are roughly equivalent as shown in Fig.~\ref{Fig4}. The results are summarized in Fig.~\ref{Fig5}. The $I_\texttt{C}$ intensities remain stable with increasing field until a threshold value of $\sim$ 0.1 T. Then they rapidly decrease as the field is increased up to $\sim$ 0.17 T ($\mu_0H \sim 0.01$ meV) (Figs.~\ref{Fig5}a, c, d), accompanied by an obvious shift of the $I_\texttt{C}$ \textbf{Q}-position from $H = 1.436(1)$ to 1.442(1) (Figs.~\ref{Fig5}b and \ref{Fig6}a). The $L$ position keeps constant (Fig.~\ref{Fig5}c). In contrast, there are no corresponding changes in the intensities or positions of $I_\pm$ (Figs.~\ref{Fig5}c and \ref{Fig6}b) or the underlying lattice, represented by the (2, 0, 0) nuclear Bragg reflection as shown in Fig.~\ref{Fig5}b. This indicates that the $I_\pm$ peaks are connected relatively closer with the underlying lattice, while the $I_\texttt{C}$ peaks appear to be relatively independent. This sharp contrast implies that the electronic origins of the two sets of peaks are completely different in nature.

\begin{figure*}[!ht]
\centering \includegraphics[width = 0.78\textwidth] {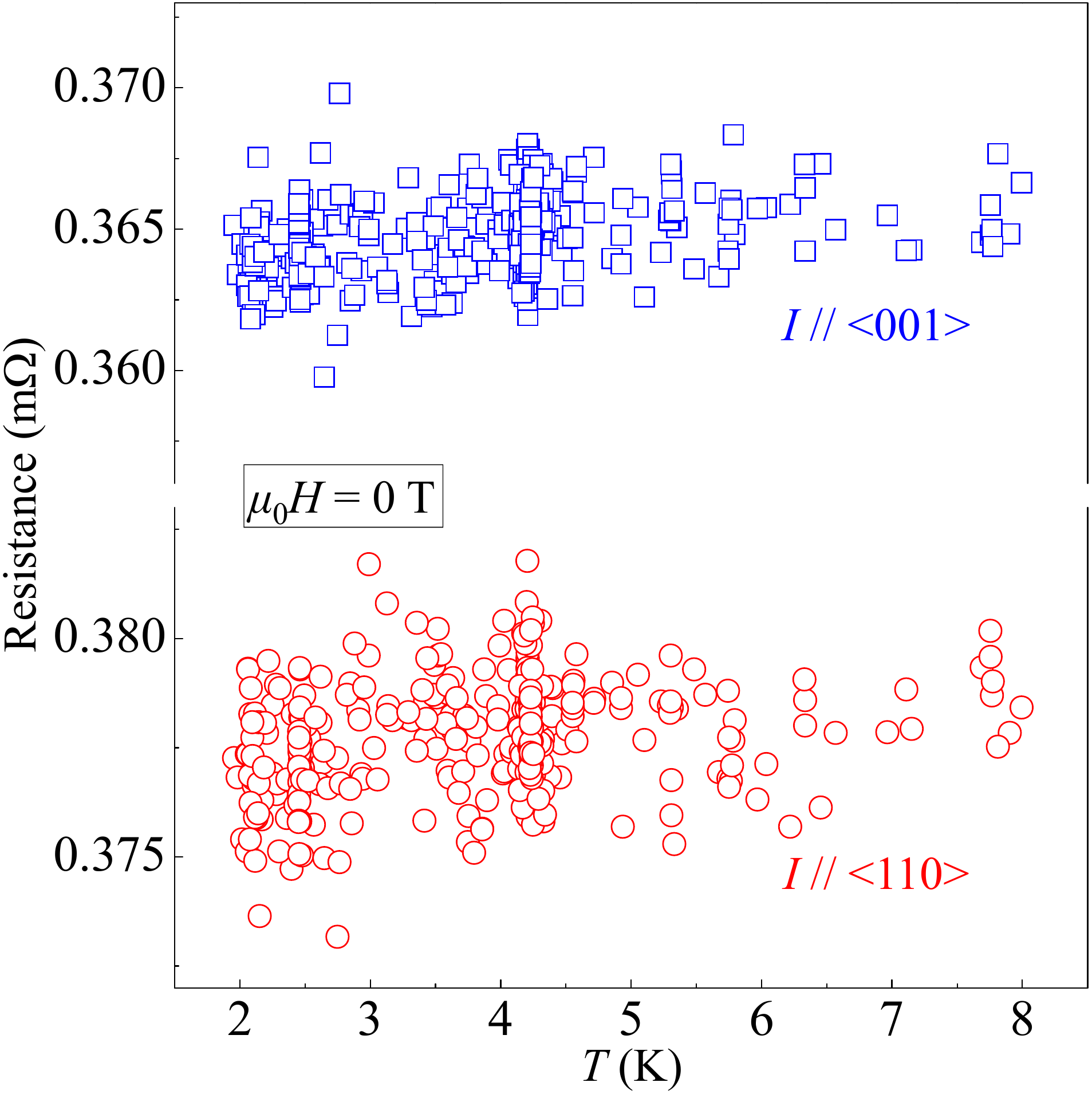}
\caption{
\textbf{Temperature variation of the resistance as measured at zero field.}
With dc current \emph{I} parallel to the $<$001$>$ axis (up) and along the $<$110$>$ zone (down). We note that the threshold of magnetic field for weakening the SDW state is so small (Fig.~\ref{Fig5}) that the commercial PPMS is unable to detect the potential weak magnetoresistance effect. The SDW state observed here represents more or less a new type of spin state since its threshold field is extremely small. Therefore, we propose that ErPd$_2$Si$_2$ is a SDW conductor.
}
\label{Fig7}
\end{figure*}

\textbf{Resistivity measurements.} The temperature-dependent resistivity along the $<$001$>$ and $<$110$>$ directions at zero field is shown in Fig.~\ref{Fig7}. Upon cooling, while entering the magnetic regimes of III, II and I as shown in Fig.~\ref{Fig4}, there is no appreciable response of resistance to the formation of the magnetic states consistent with Ref. \cite{Sampa2008}.

\textbf{Discussion}

Our data show evidence for different electronic origins of the magnetic peaks. We observe that the $I_\pm$ peaks persist down to the lowest temperature ($\sim$ 1.7 K) in this study, and they do not change position with field (Fig.~\ref{Fig5}c), and their intensities are constant below the threshold field (Fig.~\ref{Fig5}d). This is consistent with a ground-state AFM structure due to localized moments from the 4\emph{f} electrons that are tightly bound within the Er atoms.

We believe that the $I_\texttt{C}$ peaks are due to a SDW state from the following observations:
(1) The $I_\texttt{C}$ peaks are suppressed with a relatively weak magnetic field of $\sim$ 0.1 T ($\sim$ 0.006 meV), showing a fragile electronic instability that closely resembles the behaviour of CDWs driven by the electron-phonon interactions resulting from a lattice distortion \cite{Gruener1988, Johannes2008}. It is stressed that the tetragonal symmetry of the crystal lattice is reserved in the whole studied temperature range, i.e., there is no any appreciable structural phase transition. In addition, our neutron polarization analysis confirms that no such CDWs exist in ErPd$_2$Si$_2$ due to the absence of NSF scattering at the incommensurate position;
(2) The peak positions shift with a change of the applied field, independently of the underlying lattice (Fig.~\ref{Fig5}b), behaving more like a fragile Wigner-electron-crystal \cite{Wigner1934}. This indicates that the electrons responsible for the magnetic order are able to move easily in the crystal like those freely-distributed for a metallic bonding;
(3) It is unlikely that localized 4\emph{f} electrons in ErPd$_2$Si$_2$ would form two competing long-range magnetic states with the same moment direction within one pure tetragonal phase in the absence of magnetic contributions from the Pd and Si ions;
(4) The resistivity of single-crystal ErPd$_2$Si$_2$ (Fig.~\ref{Fig7}) displays no appreciable response to the appearance of $I_\texttt{C}$ peaks.
(5) The $I_\texttt{C}$ intensity is not restored on removing the magnetic field, but indeed is recovered if the temperature is raised to 10 K (above  $T_\texttt{N}$) for $\sim$ 3 mins and then cooled to 3 K.  By contrast, the $I_\pm$ intensity not only remains on the application of a field, it increases after removing the field. This clearly indicates that $I_\pm$ and $I_\texttt{C}$ have different electron origins.
Combining these five points leads us to propose that the $I_\texttt{C}$ peaks are due to a SDW state, which is a consequence of the weak (in view of the small threshold field) electron-electron interactions \cite{Gruener1994} mediated by the 4\emph{f} moments \cite{Kulikov1987}, plausibly associated with the itinerant conduction electrons near the Fermi level.

Figures~\ref{Fig3} and~\ref{Fig4} show clearly the coexistence and competition between the two modulated orders, which can be explained through a mechanism where conduction electrons contribute to both a SDW state, through Fermi surface nestings and \emph{s}-\emph{f} couplings \cite{Kulikov1987}, and to mediating the dominant RKKY interactions between localized 4\emph{f} moments \cite{Jensen1991}. These two functions are in general competitive, leading to a breakdown of the SDW state as the local-moment magnetism starts to fully develop below $\sim$ 2.9 K. A similar breakdown also occurs in the field dependence, where the fragile SDW state is suppressed above $\sim$ 0.17 T. Once above the field threshold, those conduction electrons that contribute to the SDW state then decouple and participate in the RKKY exchanges. The net result is that the $I_\texttt{C}$ \textbf{Q}-positions change, and their intensities are suppressed and unable to regain the original values when the field is removed (Figs.~\ref{Fig5}b, \ref{Fig5}d, \ref{Fig6}c and \ref{Fig6}d), whereas the $I_\pm$ intensities grow with increasing field above $\sim$ 0.17 T. Most importantly, the $I_\pm$ intensities almost linearly increase while decreasing the strength of magnetic field. The sharp suppression of the SDW state (Figs.~\ref{Fig5}a, c and d) at the threshold magnetic field is ascribed to the weak \emph{s}-\emph{f} couplings. The resistivity (Fig.~\ref{Fig7}) is insensitive to the change of the two spin states because the 4\emph{f} electrons responsible for the dominant magnetic state are localized and seldom participate in the electrical conductivity. The collectively-organized conduction electrons responsible for the weakly-pinned low-energy SDW state tend to be delocalized due to the extremely small threshold magnetic field and begin to carry current in a small potential difference.

To summarize, we have shown that two incommensurate magnetic states exist in a ErPd$_2$Si$_2$ single crystal, with different modulations propagating at \textbf{Q}$_{\pm}$ = ($H \pm 0.557(1)$, 0, $L \pm 0.150(1)$) and \textbf{Q}$_\texttt{C}$ = ($H \pm 0.564(1)$, 0, $L$), respectively, but possessing the same moment direction. We have established their temperature and magnetic-field dependent phase diagrams. We show that both states not only co-exist at $\sim3.2-4.2$ K, but they are also in competition at $\sim2.9-3.2$ K. One magnetic state correlates with the underlying lattice insofar as both are independent of the applied magnetic field in this study. In contrast, the other magnetic state is so delicate that a modest magnetic field of $\sim$ 0.17 T can suppress it, and it is not recovered on releasing the field. We thus propose that the two states have different electronic origins. One state corresponds to the localized unpaired 4$f^n$ electrons associated with the Er atoms, mediated mostly by the RKKY interactions, whereas the second one is a SDW state stemming from the conduction electrons, derived probably by the Fermi-surface nesting and/or their coupling to the local moments. We thus propose that ErPd$_2$Si$_2$ represents a prototypical model system for simultaneously studying the interesting behaviors of itinerant and localized moments well-separated in one compound.

\textbf{Methods}

\textbf{Resistivity measurements and in-house X-ray powder diffraction.} The growth and in-house characterizations of ErPd$_2$Si$_2$ single crystals were previously reported \cite{Sampa2008, Cao2013}. The electrical resistivity of a bar-shaped single crystal by standard dc four-probe technique was measured on a commercial physical property measurement system (PPMS), and a powder X-ray diffraction study was performed on an in-house diffractometer with a 2$\theta$ step size of 0.005$^\circ$ at 300 K, employing the copper $K_{\alpha 1}$ = 1.5406(9) {\AA} radiation. The powder diffraction data were analyzed by the Fullprof Suite \cite{Fullprof}.

\textbf{Crystal quality and alignment.} We selected a single piece with a mass of $\sim$ 1.12 g for the neutron scattering studies, which was oriented in the (\emph{H}, 0, \emph{L}) scattering plane of the tetragonal symmetry with the neutron Laue diffractometer, OrientExpress \cite{Ouladdiaf2006}, at the Institut Laue-Langevin (ILL), Grenoble, France. The mosaic of this single crystal is 0.48(1)$^\texttt{o}$ full width at half maximum for the nuclear (2, 0, 0) Bragg reflection at $\sim$ 230 K. Throughout this paper, the wave vector \textbf{Q}$_{(HKL)}$ ({\AA}$^{-1}$) = ($\textbf{Q}_H$, $\textbf{Q}_K$, $\textbf{Q}_L$) is defined through (\emph{H}, \emph{K}, \emph{L}) = ($\frac{a}{2\pi}Q_H$, $\frac{b}{2\pi}Q_K$, $\frac{c}{2\pi}Q_L$) quoted in units of r.l.u., where \emph{a}, \emph{b}, and \emph{c} are the relevant lattice parameters referring to the tetragonal unit cell.

\textbf{Neutron polarization analysis.} Uniaxial longitudinal neutron polarization analysis was performed at the D7 (ILL) diffractometer with incident vertically-polarized neutron spins with $\lambda$ = 4.8 {\AA}. In this polarization setup, the NSF channel collects neutron scattering intensities from the nuclear Bragg reflections and the out-of-plane magnetic moments along the $<$010$>$ direction, while the SF channel records the magnetic intensity from the in-plane moments that are perpendicular to both the \emph{b} axis and the scattering vector $\hat{\textbf{Q}}$ \cite{Stewart2009}.

\textbf{Unpolarized neutron scattering.} Elastic neutron scattering measurements were carried out on the same oriented sample at the IN12 (ILL) cold triple-axis spectrometer with a vertical moderate magnetic field (up to 3.5 T) parallel to the [010] (or [100]) direction and fixed final energy at 10.03 meV. The single crystal was top-loaded, and the beam collimation throughout the experiment was kept at open-30$'$-sample-40$'$-open.

\textbf{Acknowledgements}

This work at RWTH Aachen University and J$\ddot{\texttt{u}}$lich Centre for Neutron Science JCNS Outstation at ILL was funded by the BMBF under contract No. 05K10PA3.
C.D.C at NPU (Xi’an) acknowledges supports from National Basic Research Program of China (No. 2012CB821404) and National Natural Science Foundation of China (No. 51471135).
W.L at IFW (Dresden) acknowledges support from DFG through project SFB 463.
H.F.L is grateful to Bachir Ouladdiaf at ILL for his helpful assistance in the neutron Laue experiments, and Thomas Br$\ddot{\texttt{u}}$ckel at JCNS for helpful comments and discussions and a critical reading of the manuscript.

\textbf{Author contributions}

C.D.C. and W.L. grew the single crystals.
H.F.L. characterized the single crystal measured in this study.
A.W. and H.F.L. performed the D7 experiments.
W.S., K.S. and H.F.L. performed the IN12 experiments.
H.F.L., C.D.C., A.W., W.S., K.S., B.Y.H., L.P.R., C.Z., P.M., W.L. and G.R. discussed and analyzed the results.
H.F.L. and A.W. wrote the main manuscript text.
W.L., Th.B., A.W., C.D.C., K.S. and G.R. commented on the manuscript and all authors reviewed the paper.
H.F.L. conceived and directed the project.

\textbf{Additional information}

\textbf{Competing financial interests:} The authors declare no competing financial interests.

\textbf{How to cite this article:}

\textbf{License:} This work is licensed under a Creative Commons Attribution 4.0 International License. The images or other third party material in this article are included in the article’s Creative Commons license, unless indicated otherwise in the credit line; if the material is not included under the Creative Commons license, users will need to obtain permission from the license holder in order to reproduce the material. To view a copy of this license, visit http://creativecommons.org/licenses/by/4.0/

Correspondence and requests for materials should be addressed to H.F.L.

\newpage

\begin{table*}[!ht]
\begin{minipage} {0.64\textwidth}
\caption{Summary of the refined room-temperature structural parameters of single-crystal ErPd$_2$Si$_2$ from X-ray powder diffraction.}
\end{minipage}
\begin{minipage} {0.92\textwidth}
\label{XRPD}
\begin{tabular} {llllllllllllllllllllllllllllllllllllllllllllll}
\hline
\hline
\multicolumn{46}{c} {Pulverized single-crystal ErPd$_2$Si$_2$}  \\*
\hline
\multicolumn{7}{l} {Structure}                          &&&\vline&&&&  \multicolumn{33}{c} {Tetragonal (ThCr$_2$Si$_2$-type), $I4/mmm$} \\*
\multicolumn{7}{l} {$a, b (= a), c$ ({\AA})}            &&&\vline&&&&  & \multicolumn{8}{c} {4.0987(1)} &&&&&& \multicolumn{8}{c} {4.0987(1)} & \multicolumn{8}{c} {9.8762(1)} \\*
\multicolumn{7}{l} {$\alpha, \beta, \gamma$ $(^\circ)$} &&&\vline&&&&  & \multicolumn{8}{c} {90}        &&&&&& \multicolumn{8}{c} {90}        & \multicolumn{8}{c} {90} \\*
\hline
Atom &&&&&&&&&\vline&&&&    Site      &&&&&&&& \emph{x} &&&&&&&& \emph{y} &&&&&&&& \emph{z}   &&&&&&&& $B$ ({\AA}$^2$) \\*
Er   &&&&&&&&&\vline&&&&  2\emph{a}   &&&&&&&& 0        &&&&&&&&  0       &&&&&&&& 0          &&&&&&&& 3.06(3)  \\*
Pd   &&&&&&&&&\vline&&&&  4\emph{d}   &&&&&&&& 0        &&&&&&&&  0.5     &&&&&&&& 0.25       &&&&&&&& 3.53(3)  \\*
Si   &&&&&&&&&\vline&&&&  4\emph{e}   &&&&&&&& 0        &&&&&&&&  0       &&&&&&&& 0.3798(2)  &&&&&&&& 3.62(6)  \\*
\hline
\multicolumn{46}{c} {$R_p (\%)$: 3.09; $R_{wp} (\%)$: 4.41; $R_B (\%)$: 2.70; $R_F (\%)$: 2.19; $\chi^2$: 3.20.} \\*
\hline
\hline
\end{tabular}
\end{minipage}
\end{table*}

\end{document}